\documentclass{article}

\usepackage{scicite}
\usepackage{amsmath}
\usepackage{graphicx}
\usepackage{color}

\usepackage{times}

\title{Fifth-order susceptibility unveils growth of thermodynamic amorphous order in glass-formers}

\author{%
S. Albert$^1$, Th. Bauer$^{2, \dagger}$, M. Michl$^2$, G. Biroli$^{3,4}$, J.-P. Bouchaud$^5$, \\
A. Loidl$^2$, P. Lunkenheimer$^2$, R. Tourbot$^1$, \\
C. Wiertel-Gasquet$^1$, F. Ladieu$^{1 \ast}$\\
\\
\normalsize{$^{1}$SPEC, CEA, CNRS,  Universit\'e Paris-Saclay,}\\
\normalsize{CEA Saclay Bat $772$, 91191 Gif-sur-Yvette Cedex, France,}\\
\normalsize{$^{2}$Experimental Physics V, Center for Electronic Correlations and Magnetism,}\\
\normalsize{University of Augsburg, 86159 Augsburg, Germany,}\\
\normalsize{$^{\dagger}$ Present address: Institute for Machine Tools and Industrial Management,}\\
\normalsize{Technical University of Munich, 85748 Garching, Germany,}\\ 
\normalsize{$^{3}$IPhT, CEA, CNRS, Universit\'e Paris-Saclay,}\\
\normalsize{CEA Saclay Bat $774$, 91191 Gif-sur-Yvette Cedex, France,}\\
\normalsize{$^{4}$LPS, Ecole Normale Sup\'erieure, 24 rue Lhomond, 75231 Paris Cedex 05, France,}\\
\normalsize{$^{5}$Capital Fund Management, 23 rue de l'Universit\'e, 75007 Paris, France}\\
\\
\normalsize{$^\ast$Corresponding author:  francois.ladieu@cea.fr.}%
}

\date{}
\begin{document} 

\maketitle 

\begin{quote}\textbf{
Glasses are ubiquitous in daily life and technology. However the microscopic mechanisms generating this state of matter remain subject to debate: Glasses are considered either  
as merely hyper-viscous liquids or as resulting from a genuine thermodynamic phase transition towards a rigid state. We show that third and fifth order susceptibilities 
provide a definite answer to this longstanding controversy. Performing the corresponding high-precision nonlinear dielectric experiments for supercooled glycerol and propylene carbonate, we find strong support 
for theories based upon thermodynamic amorphous order. Moreover, when lowering temperature, we find that the growing transient domains are compact -- that is their fractal dimension $d_f  =  3$. The 
glass transition may thus represent a class of critical phenomena different from canonical second-order phase transitions for which $d_f < 3$.%
}\end{quote}

The glassy state of matter, despite is omnipresence in nature and technology \cite{Edi12}, continues to be one of the most puzzling riddles in condensed-matter physics  \cite{Edi12,Ber11}: For all practical purposes, glasses are rigid like crystals but they lack any long-range order. Some theories describe glasses as kinetically constrained liquids \cite{kcm}, becoming so viscous below the glass transition that they seem effectively rigid. By contrast, other theories \cite{Wolynes,Gilles} are built on the existence of an underlying thermodynamic phase transition to a state where the molecules are frozen in well defined,
 yet disordered positions.  This so-called "amorphous order" cannot be revealed by canonical static correlation functions, but rather by new kinds of correlations 
 [i.e. point-to-set correlations or other measures of local order \cite{BBCGV,TarjusMossa}] that have now been detected in recent numerical 
 simulations \cite{BBCGV,Berthieretal,Royalletal}.  In these theories, thermodynamic correlations lock together the fluctuations and response 
 of the molecules, which collectively rearrange over some
length-scale $\ell$, ultimately leading to rigidity. In this thermodynamic scenario, $\ell$ is proportional to a power of  $\ln(\tau_{\alpha}/\tau_{0})$ where
$\tau_{\alpha}$ is the structural relaxation time and $\tau_0$ is the microscopic time-scale, generally smaller than $1$ps \cite{Wolynes,Gilles}. Because equilibrium measurements require a time larger than $\tau_{\alpha}$, 
they cannot be performed in
 the range where $\ell$ is very large since this would require exponentially long times. This limitation is essentially why the true nature of glasses is still a matter of intense debate.

Here, we propose a pioneering strategy to unveil the existence of a thermodynamic length $\ell$ that grows upon cooling. 
Instead of only varying the temperature $T$, we also vary the non-linear order $k$ of the response 
of supercooled liquids. This is motivated by a general, although rarely considered \cite{Kim00}, property of critical points: At a second order critical temperature $T_c$, 
the linear susceptiblity $\chi_{1}$ associated to the order parameter is not the only diverging response. 
As a function of temperature, all the higher order responses $\chi_{2m+1}$ with $m \ge 1$ 
diverge even faster than $\chi_1$ itself. This comes from the fact that the divergencies of all the $\chi_{2m+1}$ have the same origin, namely the divergence  
of the length $\ell$. By using the appropriate scaling theory, it can be shown that the larger $m$, the stronger the divergence in temperature. As theoretically shown below, 
transposing this idea to glasses requires taking into account that the putative ``amorphous'' or hidden order in supercooled liquids \cite{BBJCP,BBCGV} 
is not reflected in $\chi_1$ itself, but only in higher-order response functions $\chi_{2m+1}$ with $m \ge 1$. This idea is indeed supported by previous measurements and analyses of 
the third order susceptibility $\chi_3$ \cite{Cra10,Bau13,Bru12,Buc14,Cas15}. 
We report results on the fifth order 
susceptibility $\chi_5(T)$ and compare them to $\chi_3(T)$ in two canonical glass forming liquids, glycerol and propylene carbonate. 
If critical phenomena really play a key role for the glass transition, $\chi_5$ should increase much faster than $\chi_3$ as the liquid becomes more viscous.
 
This scenario can be understood by means of a theoretical argument based on previous work \cite{Bou05} and further detailed in \cite{SI}. Suppose 
that $N_{\text{corr}} = (\ell/a)^{d_f}$ molecules are amorphously ordered over the length-scale $\ell$, where $a$ is the molecular size 
and $d_f$ is the fractal dimension of the ordered clusters. This implies that their dipoles, oriented in apparently random positions, 
are essentially locked 
together during a time $\tau_\alpha$. We expect that in the presence of an external electric field $E$ oscillating at frequency $\omega \ge \tau_\alpha^{-1}$, the dipolar degrees of freedom of these molecules contribute to the polarisation per unit volume as 
 
\begin{equation} \label{scaling}
p = \mu_{dip} \frac{\sqrt{(\ell /a)^{d_f}}}{(\ell /a)^d} F\left(\frac{\mu_{dip} E \sqrt{(\ell /a)^{d_f}}}{kT}\right)
\end{equation}
where $\mu_{dip}$ is an elementary dipole moment, $F$ is a scaling function such that $F(-x)=-F(x)$, and $d=3$ the dimension of space.
This states that randomly locked dipoles have an overall moment $\sim \sqrt{N_{\text{corr}}}$, and that we should compare the energy of this ``super-dipole'' in a field to the thermal energy. Equation \ref{scaling} is  motivated by general arguments involving multi-point correlation functions through which $d_f$ can be given a precise meaning \cite{SI} and is fully justified when $\ell$ diverges, in particular in the vicinity of a critical point such as the Mode-Coupling transition or the spin-glass transition. In the latter case, Eq. \ref{scaling} is in fact equivalent to the scaling arguments of \cite{LL}, provided one performs the suitable mapping between the magnetic formalism of \cite{LL} and ours.

Expanding Eq. \ref{scaling} in powers of $E$, we find the ``glassy'' contribution to
$p$:
\begin{eqnarray} \label{eq2}
\frac{p}{\mu_{dip}} & = & F'(0) \left(\frac{\ell}{a} \right)^{d_f - d} \left(\frac{\mu_{dip} E}{kT}\right) +
\frac{1}{3!} F^{(3)}(0) \left(\frac{\ell}{a} \right)^{2d_f - d} \left(\frac{\mu_{dip} E}{kT}\right)^3 + \nonumber  \\
\ \ &\ & + \frac{1}{5!}F^{(5)}(0) \left(\frac{\ell}{a} \right)^{3d_f - d} \left(\frac{\mu_{dip} E}{kT}\right)^5 + \dots
\end{eqnarray}
Because $d_f$ must be less or equal to $d$, we find that the first term, contributing to the usual linear dielectric constant $\chi_1(\omega)$,
cannot grow as $\ell$ increases. This simple theoretical argument explains why we do not expect spatial glassy correlations to show up in $\chi_1(\omega)$. The second term,
contributing to the third-order dielectric constant, does grow with $\ell$ provided $d_f > d/2$. Although $d_f <d$ close to a standard second order critical point \cite{Coniglio} such as the spin-glass
transition, several theories suggest \cite{Wolynes,Gilles,WolynesSchmalian,reviewRFOT} that ordered domains are compact ($d_f=d$), in
which case $\left(\frac{\ell}{a} \right)^{2d_f - d} = \left(\frac{\ell}{a} \right)^{d} = N_{\text{corr}}$, as assumed in our previous papers \cite{Bou05,Lad12}. 
The third term of Eq. \ref{eq2} reveals that the fifth-order susceptibility $\chi_5(\omega)$ should diverge as $\ell^{3d_f - d}$. Therefore, the joint measurement of
$\chi_3(\omega)$ and $\chi_5(\omega)$ provides a direct way to estimate $d_f$ experimentally, through the following relation:
\begin{equation} \label{def-tau}
|\chi_5| \propto |\chi_3|^{\mu(d_f)}; \qquad \mu(d_f)=\frac{3d_f-d}{2 d_f-d}; \qquad d_f(\mu)=d \, \frac{\mu - 1}{2 \mu - 3}
\end{equation}
where the exponent $\mu(d_f)$ is equal to $2$ when the dynamically correlated regions are compact ($d_f=d$), and is higher otherwise. We predict two key results that can obtained from  $\chi_5$
and $\chi_3$ susceptibility measurements. First, if amorphous order increases approaching the transition, the frequency dependence should be more anomalous 
[i.e. more humped \cite{SI}] for $\chi_5(\omega)$ than for $\chi_3(\omega)$. Second, the growth of $\chi_5$ should be much stronger than that of $\chi_3$ when lowering the
 temperature, following $\chi_5\sim \chi_3^2$ if we assume compact amorphous domains.  
Our work provides experimental evidence that these predictions indeed hold and
suggests that the glass transition represents a new type of critical phenomenon with growing length and time scales but with $d_f=d$, in contrast to the
spin-glass transition that instead displays \cite{LL} canonical critical behavior with $d_f \approx 2.35$.

We measured $\chi_5(\omega)$ in two canonical glass formers, glycerol and propylene carbonate, by applying a field of amplitude $E$ and of
frequency $f= \omega /(2 \pi)$, see \cite{SI}.
The fifth-order response is $\propto \chi_5 E^5$, and orders of magnitude smaller than the cubic and
linear ones, given by $\propto \chi_3 E^3$ and $\propto \chi_1 E$ respectively. We avoided any contributions of $\chi_3$ and of $\chi_1$ by measuring the signal at $5 \omega$, 
which only contains the component $\chi_5^{(5)}$ of the fifth order susceptibility \cite{SI}. We measured $\chi_5^{(5)}$ with two independent setups due to the very small amplitude, 
optimized along complementary strategies. One setup (in Augsburg) was designed to achieve the highest possible field (reaching $78$ MV/m). We optimized sensitivity with a 
differential technique using two samples of different thicknesses in the other setup (Saclay, see Fig. S1 of \cite{SI}), which required lower fields (up to $26$ MV/m).

We have obtained the values of $\vert \chi_5^{(5)}(\omega)\vert $ for glycerol at various frequencies and temperatures by using the two aforementioned techniques (Fig. 1A). 
A clear peak arises for a given $T$ in $\vert \chi_5^{(5)}(\omega)\vert$ for a frequency $f_{\text{peak}} \simeq 0.22 f_{\alpha}$ where the $\alpha$-relaxation
frequency $f_{\alpha}$, defined by the peak of the out-of-phase linear susceptibility, is indicated by arrows in Fig. 1A. Even though the data were determined by two independent setups, the overall agreement is remarkable (Fig. 1B). The most accurate comparison is possible at $204$K where $f_{peak}$ is well inside the frequency range accessible by the two setups. The two spectra  at $204$K coincide on the low frequency side of the peak \cite{SI}. On the other side of the peak, a discrepancy between the
two sets of data progressively increases with frequency, reaching a constant factor of $4$ at the highest frequencies (Fig. 1B). Apart from the value of the electric field, the main
difference between the two experiments is the number of applied field cycles $n$. The Saclay setup measured the stationary responses ($n \to \infty$), whereas $n$ remained finite in the Augsburg setup [similarly to \cite{Ric06}], ranging from $n=2$ at the lowest frequencies to $n \propto f$ at the
highest frequencies. The two setups give the same results for $\chi_5^{(5)}$ because at sufficiently low values
 of $f/f_\alpha$, the response of the supercooled liquid is likely to instantaneously follow the field. By contrast, at higher frequencies $\omega \ge \tau_{\alpha}^{-1}$ the finite cycle number may play a role, making a quantitative treatment of this effect difficult \cite{SI}. Our further analysis relies on the behavior
 of the peaks of $\chi_5^{(5)}$, and more precisely on their relative evolution with temperature, which reasonably agrees in the two setups (see below).
	
The qualitative features of $\vert \chi_5^{(5)} (\omega) \vert$ (Fig. 1A-B) are reminiscent of those of the third harmonic
cubic susceptibility $\vert \chi_3^{(3)} (\omega)\vert$, \cite{Cra10,Bau13}. Both quantities exhibit a humped shape, with
a peak located at the same frequency $f_{\text{peak}} \approx 0.22 f_{\alpha}$, as well as a strong increase of the height of the peak as the temperature is decreased.
These two distinctive features are important since they are specific signatures of glassy dynamical correlations \cite{Bou05}, in contrast
to trivial systems without correlations \cite{Cof76}. In this
case, the modulus of all higher-order non linear susceptibilities monotonously decreases with frequency \cite{Cof76,SI}.

To quantitatively compare the frequency dependency of the susceptibilities $\chi_k^{(k)}$ of order $k$, 
we plotted \\ $\vert \chi_k^{(k)}(f/f_{\alpha})/ \chi_k^{(k)}(0)\vert$ of glycerol for $k=5$, $3$, and $1$ ($\chi_1^{(1)}$ is the linear susceptibility noted $\chi_1$ above) (Fig. 1C). 
The peak amplitude for $k=5$ is strongly 
enhanced compared to $k=3$ -- that is the higher the nonlinear order $k$,
the more anomalous the frequency dependence (Fig. 1C and Figs. S2-S3 of \cite{SI}). This behavior is a decisive result and fully consistent with our scaling analysis. For archetypical glass formers, we can always fit the linear
susceptibility by assuming a sum of Debye relaxations where $\chi_{1 \text{, Debye}} \propto 1/(1-i \omega \tau)$. We do this by choosing a suitable
distribution $\cal{G}(\tau)$ of relaxation times $\tau$ \cite{Blo03} caused by dynamical heterogeneities.
Because the trivial response discussed above also obeys $\chi_{1 \text{, trivial}} \propto 1/(1-i \omega \tau)$, we have used \cite{SI,Cof76} the same
distribution ${\cal {G}}(\tau)$ to calculate the trivial responses $\chi^{(k)}_{k \text{, trivial}}$ for $k=3$ and $5$.
For a given $k>1$, a large difference exists between the experimental spectrum of
 $\vert \chi^{(k)}_k(f/f_{\alpha})/ \chi^{(k)}_k(0) \vert$ and its trivial counterpart (Fig 1C), which we can ascribe to correlation-induced effects. For $k=1$
 the experimental data agree with the trivial response [convoluted with ${\cal {G}}(\tau)$], consistent with the theoretical arguments stating that glassy correlations do not change the linear
 response \cite{Bou05}. For $k=3$ and $5$ the difference to the trivial response increases, being much more important for $k=5$ where it exceeds one
  order of magnitude. This quantitatively supports the scaling prediction obtained assuming that collective effects due to the growth of
  amorphous order play a key role in supercooled liquids.

We measured $\vert \chi_5^{(5)}(\omega) \vert$ at five different temperatures for propylene carbonate (Fig. 2). Propylene carbonate 
differs from glycerol in that its fragility \cite{Deb01,Boh93} ${\text{m}} \propto [\partial \log(\tau_{\alpha})/\partial (1/T)]_{T_g}$ ($T_g$ denotes the glass transition temperature) is 
twice as large and it has Van der Waals bonding, in contrast to hydrogen bonding. Despite these differences, the anomalous hump-like features of glycerol 
(Fig. 1A) are also observed in propylene carbonate (Fig. 2). We expect this behavior from our scaling framework, which relies on the predominant
role of collective dynamical effects in supercooled liquids. The presence of similar anomalous features in two very different glass formers suggests they only weakly depend on the 
specific microscopic properties of the material.

To elicit the temperature dependence of collective effects, we introduced dimensionless quantities related to $\chi_3^{(3)}$ and $\chi_5^{(5)}$:
\begin{equation}
X_3^{(3)}\equiv \frac{k_B T}{\epsilon_0 \Delta \chi_1^2 a^3} \chi_3^{(3)}, \qquad
X_5^{(5)} \equiv \frac{(k_B T)^2 }{\epsilon_0^2 \Delta \chi_1^3 a^6} \chi_5^{(5)}
\label{eqX5}
\end{equation}
where $\epsilon_0$ is the permittivity of free space, $\Delta \chi_1 = \chi_1(0) -\chi_1(\infty)$ is the dielectric strength,
$a^3$ is the molecular volume and $k_B$ is the Boltzmann constant. The main advantage of these dimensionless non-linear susceptibilities is that
in the trivial case of an ideal gas of dipoles, both $X_{3 \text{, trivial}}^{(3)}$ and $X_{5 \text{, trivial}}^{(5)}$ are
independent of temperature when plotted versus scaled frequency \cite{Cof76,SI}. Hence, we ascribe their experimental variation to the non-trivial dynamical
correlations in the supercooled liquid \cite{Bou05,Lad12}. This interpretation is strongly supported by previous findings \cite{Cra10,Bru12,Lad12,Bau13} where the temperature dependence of
$\vert X_{3}^{(3)} \vert$ was studied at various values of $f/f_{\alpha}$. Close to and above its peak frequency, $\vert X_{3}^{(3)} \vert$ was found to strongly vary in
temperature, contrary to the low-frequency plateau region ($f/f_{\alpha} \le 0.05$) where $\vert X_{3}^{(3)} \vert$ no longer depends on temperature. This
low frequency region corresponds to time scales much longer than $\tau_{\alpha}$ where the liquid flow destroys glassy correlations, making each molecule effectively
 independent of others, and yielding a dielectric response close to the aforementioned trivial case. This is why, to determine the temperature evolution of the glassy dynamical correlations,
 we focused on the region of the peak of $\vert \chi_5^{(5)} \vert$. For each of the two liquids, this peak appears at the very same 
 frequency $f_{\text{peak}}$ as in $\vert \chi_3^{(3)} \vert$.

We expect the nonlinear susceptibilities to contain both a trivial contribution that would exist even for independent dipoles, and a ``singular'' contribution (i.e. diverging with $\ell$) 
as given in Eq. \ref{eq2}. We thus write:

\begin{equation}
X_{3 \text{, sing.}}^{(3)} \equiv X_{3}^{(3)} - X_{3 \text{, trivial}}^{(3)}, \qquad X_{5 \text{, sing.}}^{(5)} \equiv X_{5}^{(5)} - X_{5 \text{, trivial }}^{(5)}
\label{eqX5sing}
\end{equation}

\noindent Here the trivial contributions are calculated by assuming a set of independent Debye dipoles convoluted with the aforementioned distribution $\cal G(\tau)$ of relaxation times  \cite{SI}.
 We compared the temperature evolution
of \\ $\vert X_{5{\text{,sing.}}}^{(5)}(f_{\text{peak}}(T)) \vert$ and that of $\vert X_{3 {\text{,sing.}}}^{(3)} (f_{\text{peak}}(T)) \vert^\mu$ (Fig 3), to derive the value of the
exponent $\mu$, from which we deduce the fractal dimension $d_f$ of the dynamically correlated regions by using Eq. \ref{def-tau}.
In both glycerol and propylene carbonate, the value $\mu =2$, corresponding to compact domains of dimension $d_f=3$, is found to be consistent with
experiments (triangles in Figs 3A and B). By fitting the $T$ dependence of $\vert X_{3 {\text{,sing.}}}^{(3)}(f_{\text{peak}}(T)) \vert$ with a smooth function \cite{SI}, we found 
the hatched area corresponding to $\mu = 2.2 \pm 0.5$ in glycerol and $\mu = 1.7 \pm 0.4$ in propylene carbonate (Fig. 3). The fact that, within experimental uncertainty,
a value of $\mu \simeq 2$ is common to each of the two liquids supports a picture of amorphous compact domains mostly independent of differences at the molecular level and validates 
the correlation length-scale for our scaling analysis. Considering that the temperature interval in Fig. 3B is smaller by a factor of $2$, we note that the critical 
behavior in propylene carbonate 
is stronger than in glycerol (Fig. 3A). This suggests that the larger the fragility, the stronger the temperature dependence of the thermodynamic length $\ell$. This is easily 
understood in the scenario of \cite{Wolynes} where the critical point is the Vogel-Fulcher temperature $T_0$: In this case, equilibrium measurements can be made closer to the critical point for more fragile liquids, because the larger the fragility, the smaller the difference between $T_g$ and $T_0$. 

Our experimental results are therefore consistent with the general predictions of theories - such as the Random First Order Transition or Frustration Limited
Domains \cite{Wolynes,Gilles} - where the physical mechanism driving the glass transition is of thermodynamic origin and where some non trivial (albeit random) long-range correlations 
build up between molecules. Only in this case \cite{SI} can one have $N_{\text{corr}}$ dipolar degrees of freedom collectively responding over some length-scale $\ell$ and on time-scales
of the order of $\tau_\alpha$. If instead the glass transition is regarded as a purely dynamical phenomenon, there would not be any anomalous increase of the 
normalized peak value of the higher-order susceptibilities at all \cite{SI}.
Our results therefore severely challenge theories advocating against any thermodynamic signature and favoring purely dynamic scenarios.
Moreover, from a comparison of the higher-order susceptibilities, our results are consistent with $\chi_5 \propto \chi_3^2$. This  constitutes  
evidence for compact amorphously ordered domains, i.e. $d_f=d$, pointing towards a
non-standard nature of the glass transition, in contrast to canonical second order phase transitions for which $d_f<d$.

\noindent \textbf{ACKNOWLEDGMENTS}

\noindent We thank C. Alba-Simionesco, A. Coniglio, P.-M. D\'ejardin, G. Tarjus, and M. Tarzia for interesting discussions. This work in Saclay has been supported 
in part by ERC grant NPRGLASS, by the Labex RTRA
grant Aricover and by the Institut des Syst\`emes Complexes ISC-PIF. The work in Augsburg was supported by the Deutsche Forschungsgemeinschaft via Research Unit FOR1394.
S.A., R.T., C.W-G. and F.L. developed the Saclay's experimental setup. Th.B. and P.L. developed the experimental setup in Augsburg. Th.B. and M.M. performed the measurements and
analysis of the Augsburg data. A.L. and P.L. conceived and supervised the project in Augsburg. G.B. and J.-P.B. derived the theoretical scaling analysis; and all authors were
involved in the interpretation of the results and creation of this manuscript. Data are available as supplementary material. All authors have no competing financial interests. \\

\noindent \textbf{SUPPLEMENTARY MATERIALS}

\noindent www.sciencemag.org/content/352/6291/1308/suppl/DC1 \\
Materials and Methods\\
Figs. S1 to S3\\
References \cite{IMCT,Thi08,Mur12,Ama16,fisher,janus,Tarzia,BT,dyre,GC,Bru11b,Dej15,Dej00,Bru11}\\
Data files

\begin{figure}[t]
\centerline{
\includegraphics[keepaspectratio,width=520.8pt]{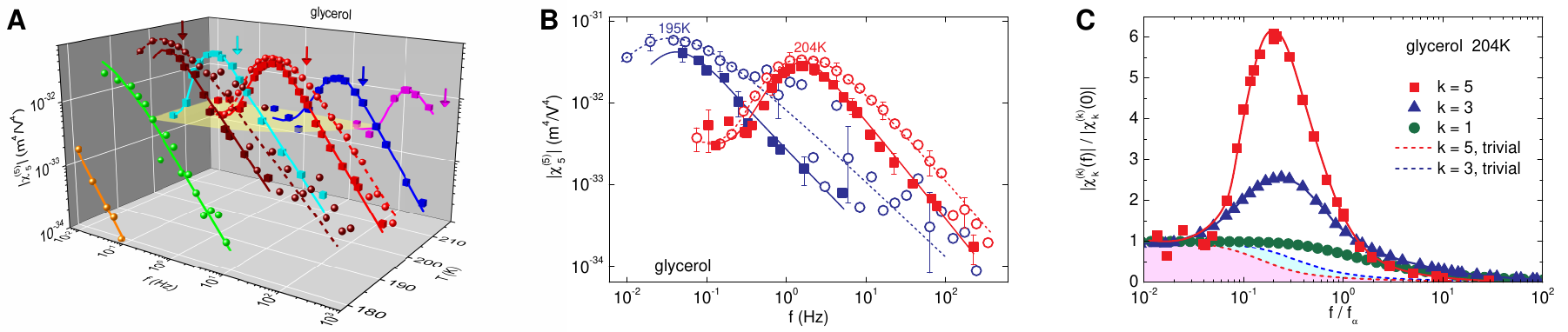}%
}
\caption{{\bf{Modulus of the fifth-order susceptibility in supercooled glycerol measured with two independent setups. (A)}} The susceptibilities $\chi_5^{(5)}$ reported here are obtained directly \cite{SI} by monitoring the response of the sample at $5 \omega$, when applying an electric field $E$ at angular frequency $\omega$. Two independent setups were used, designed either to maximize the field amplitude (Augsburg setup, spheres), or to optimize the sensitivity (Saclay setup, cubes). Lines are guides for the eyes. Errors are of the order of the scatter of neighboring data points around the lines. Both setups yield consistent results. For a given temperature $T$, $\vert \chi_5^{(5)} \vert$ has a humped shape, with a maximum occurring at the frequency $f_{\text{peak}} \simeq 0.22 f_{\alpha}$ where $f_{\alpha}$ is the relaxation frequency indicated by a colored arrow for each temperature. When decreasing $T$, the height of the hump increases strongly. The yellow plane emphasizes the fact that, for a given $T$, $\chi_5^{(5)}$ is constant for  $f \le 0.05 f_{\alpha}$. {\bf{(B)}} Projection onto the susceptibility-frequency plane of the data of Fig. 1A at $204$K and at $195$K. The agreement around and below the peak is remarkable at $204$K (see text). The relative evolution of the height of the peak is reasonably similar between $204$K and $195$K for the two setups (see Fig. 3A). {\bf{(C)}} Comparison of the fifth-order, cubic, and linear susceptibilities of glycerol [the latter is noticed $\chi_1^{(1)}$ for convenience, see \cite{SI}]. Symbols, with line to guide the eyes, are Saclay data at $204$K; the error bars are of the order of the size of the symbols for $k=5$ [except at the lowest frequencies, see \cite{SI}] and smaller for $k=3$ and $1$. The higher the order $k$, the stronger the hump of $\vert \chi_k^{(k)} \vert$: this is a key result supporting the amorphous-order scenario. The dashed lines, emphasized by colored areas, correspond to the trivial response of an ideal gas of dipoles without amorphous order. In this case $\vert \chi_k^{(k)} \vert$ decreases monotonously in frequency for any value of $k$. The higher $k$, the stronger the difference between the measured and trivial susceptibility.} 
\end{figure}
\clearpage

\begin{figure}[t]
\centerline{
\includegraphics[keepaspectratio,width=165.6pt]{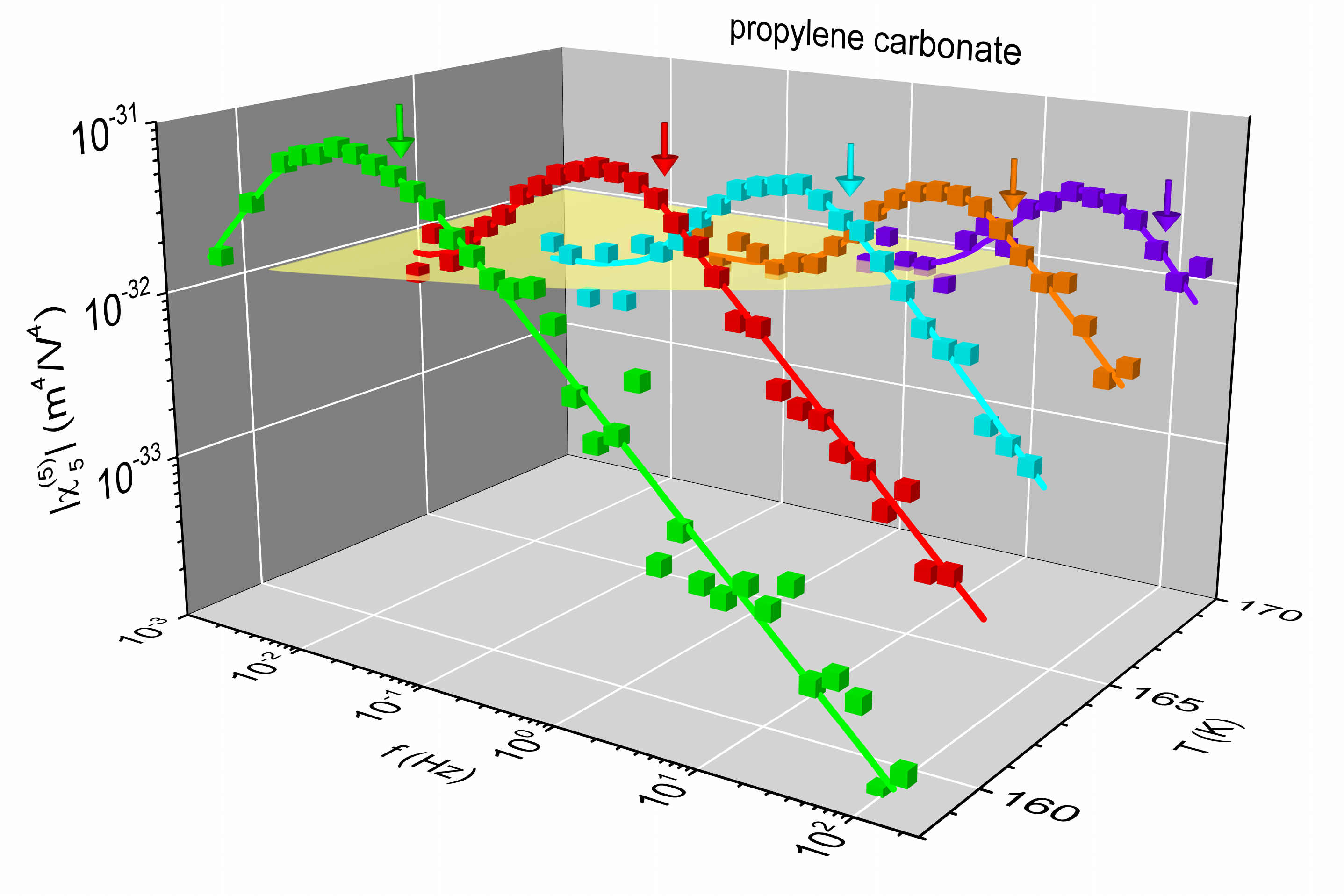}%
}
\caption{{\bf{Modulus of the fifth-order susceptibility in supercooled propylene carbonate}}. The experimental data (symbols) were obtained with the Augsburg setup. The presentation of the graph is analogous to that of Fig 1A to emphasize the similarity of the behavior of $\vert \chi_5^{(5)} \vert$ in propylene carbonate and in glycerol, even though 
these two liquids have different fragilities and different types of intermolecular interactions (van der Waals versus hydrogen bonding).}
\end{figure}

\begin{figure}[b]
\centerline{
\includegraphics[keepaspectratio,width=343.2pt]{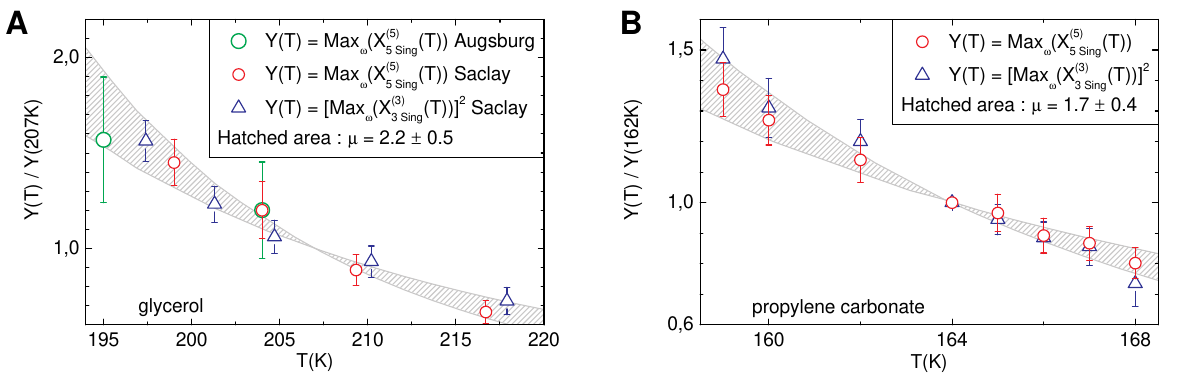}%
}
\caption{{\bf{Comparison of the temperature dependence of the singular part of the fifth order and cubic dimensionless susceptibilities at $f_{\text{peak}}$. (A)}} For glycerol, the singular part of $\vert X_k^{(k)} (f_{\text{peak}}) \vert$ for $k=3$ and $5$ is normalized to $1$ at $207$K. The value of the exponent $\mu$ is then determined by comparing  $\vert X_{5 \text{,sing}}^{(5)} (f_{\text{peak}}) \vert$ to $\vert X_{3 \text{,sing}}^{(3)} (f_{\text{peak}}) \vert^\mu$: the symbols for $k=3$ correspond to $\mu = 2$ and the hatched area to the interval corresponding to the error bar given for $\mu$ \cite{SI}. The two Augsburg data points for $X_5^{(5)}$ have been added on the graph by scaling to the Saclay point at $204$K: the Augsburg point at $195$K is reasonably well within the hatched area, which shows that the relative evolution of $X_5^{(5)}$ with temperature is consistent in the two setups.  {\bf{(B)}} Same display as in  (A) but for propylene carbonate with $T=164$K as the normalization temperature and the symbols for $k=3$ corresponding to $\mu = 2$.}
\end{figure}

\end{document}